# Preparation of $CH_3NH_3PbI_3$ thin films with tens of micrometer scale at high temperature

Hao Zhang [1], Mian Tao[1], Baizhi Gao[1], Wei Chen[1], Qi Li [1], Qingyu Xu [1,2,*], and Shuai Dong [1,*]

**Abstract**

The fabrication of high-quality organic-inorganic hybrid halide perovskite layers is the key prerequisite for the realization of high efficient photon energy harvest and electric energy conversion in their related solar cells. In this article, we report a novel fabrication technique of $CH_3NH_3PbI_3$ layer based on high temperature chemical vapor reaction. $CH_3NH_3PbI_3$ layers have been prepared by the reaction of $PbI_2$ films which were deposited by pulsed laser deposition, with $CH_3NH_3I$ vapor at various temperatures from 160 °C to 210 °C. X-ray diffraction patterns confirm the formation of pure phase, and photoluminescence spectra show the strong peak at around 760 nm. Scanning electron microscopy images confirm the significantly increased average grain size from nearly 1 μm at low reaction temperature of 160 °C to more than 10 μm at high reaction temperature of 200 °C. The solar cells were fabricated, and short-circuit current density of 15.75 mA/cm$^2$, open-circuit voltage of 0.49 V and fill factor of 71.66% have been obtained.

[1]Department of Physics, Southeast University, Nanjing 211189, China. [2]National Laboratory of Solid State Microstructures, Nanjing University, Nanjing 210093 China.

*Corresponding authors: xuqingyu@seu.edu.cn (Q. X.), sdong@seu.edu.cn (S. D.)

# Introduction

The fast exhausted traditional fossil fuels are unable to meet the rapid increasing demands of mankind, and solar energy as the most abundant energy resource on earth can readily satisfy the needs of people if it is efficiently used.[1] In recent years, organic-inorganic hybrid halide perovskites have become the hottest light absorption materials in thin film solar cell research field, on account of their appropriate band gap, low exciton binding energy (< 10 meV), high absorption coefficient of $1.5×10^4$ $cm^{-1}$ at 550 nm, wide absorption spectrum up to 800 nm, high charge carrier mobility (ca. 66 $cm^2V^{-1}s^{-1}$), and long charge diffusion length (up to 1 μm).[2–4] The formula of these perovskite compounds is $ABX_3$ (A cations are organic like $CH_3NH_3^+$, $C_2H_5NH_3^+$ and $HC(NH_2)_2^+$; B cations are metal elements such as $Pb^{2+}$ and $Sn^{2+}$ from Group IVa; X anions are halogen elements ($I^-$, $Br^-$, $Cl^-$) ).[5,6] Perovskite materials as light absorber can be tuned either by changing the alkyl group, or metal atom and halide due to that they have wide direct band gaps.[7] Many methods have been used to prepare the organic-inorganic hybrid halide perovskite thin films, such as one-step method, sequential deposition, pulsed laser deposition (PLD), silk-screen printing, chemical vapor deposition, and so on.[6-10] The theoretical limit efficiency of these perovskite cells is around 31% and the peak acquired efficiency has quickly reached up to 22.1% since the first application of organic-inorganic hybrid halide perovskite in dye-sensitized solar cell in 2009.[11-13] The current experimental works demonstrate that grain size has an obvious impact on the films' optoelectronic properties.[16] The efficiency losses resulting from defects and traps can be significantly reduced by improving crystallinity and enlarging grain size.[17–19] Due to fewer grain boundaries, those films with large size grains can exhibit higher carrier mobility and longer electron-hole recombination time than those films with small size grains.[20,21] Thus, increasing the grain size is a key for the preparation of high-quality organic-inorganic hybrid halide perovskite thin films.[22,23] Lilliu *et al* found that annealing $CH_3NH_3PbI_3$ ($MAPbI_3$) thin films can effectively increase the grain size in $CH_3NH_3I$ (MAI) atmosphere.[24] However, $MAPbI_3$ thin films are easily decomposed at high temperature due to thermal instability.[19]

In this work, we report a novel preparation method of $MAPbI_3$ films with large grain size by reacting $PbI_2$ films in the presence of MAI vapor at high temperature.[12] Using this method, we successfully fabricated pure-phase $MAPbI_3$ thin films with grain size from a few micrometers to dozens of micrometers at reaction temperature of 160 °C-220 °C. Planar structure solar cells were fabricated with $MAPbI_3$ thin films prepared by this method and a photoelectric conversion efficiency of 5.56% with $J_{sc}$ of 15.75 mA/cm$^2$, $V_{oc}$ of 0.49 V, and FF of 71.66% were achieved.

**Results and discussion**

In our work, the thickness of the $MAPbI_3$ layer is related to thickness of the $PbI_2$ layer prepared by pulsed laser deposition (PLD) and we can change thickness of $MAPbI_3$ layer by increasing or reducing thickness of the $PbI_2$ layer by regulating the energy or pulse number of laser. The $PbI_2$ thin film and MAI powder were put into the ceramic container, as shown in Figure 1. The ceramic container was put into a tube furnace to be heated. At the bottom of container, MAI powder sublimated to produce

MAI gas, and partly might be decomposed to generate HI gas at high temperature.[26] All MAI gas produced from MAI powder was gathered by a cover plate with a square hole and reacted with the PbI$_2$ thin film on the top to prepare the MAPbI$_3$ thin film. The redundant gas can be vented through the pipe on the top. Every PbI$_2$ thin film needs at least 2 g MAI powder to maintain enough MAI atmosphere until reaction is over. The formed MAPbI$_3$ will not further react with MAI or HI vapor, and the MAI vapor can suppress the decomposition of MAPbI$_3$.

From the X-ray diffraction (XRD) pattern shown in Figure 2(a), we can see the existence of a strong peak of lead at 32° in PbI$_2$ thin film which indicates the existence of Pb in PbI$_2$ thin film prepared by PLD. However, the XRD pattern (Figure 2(b)) of the prepared MAPbI$_3$ thin film at 180 °C for 25 minutes shows that all peaks are in agreement with those of pure MAPbI$_3$, indicating that all PbI$_2$ in the film have transformed into MAPbI$_3$ without any Pb and PbI$_2$ left after the reaction in the resistance furnace with the powder of MAI at high temperature.[15,27,28] Figure 2(c) exhibits the peak of photoluminescence (PL) spectrum at around 765 nm, which accords with the characteristic of pure MAPbI$_3$ thin films.[29] The absorption edge of the prepared MAPbI$_3$ thin film reaches 800 nm and the light of wavelength less than 580 nm can be totally absorbed. As shown in inset of Figure 2(c), the as-prepared PbI$_2$ thin film appears in yellow. After the complete reaction, the color of MAPbI$_3$ thin film changes to brownish red. The prepared layer starts to transform into the perovskite structure, which is very prominent at high temperature.[19]

It has been pointed out that metallic Pb is left in the as-prepared PbI$_2$ films but pure phase MAPbI$_3$ is formed without Pb left after the high temperature reaction. To clarify the mechanism, a PbO thin film was prepared by PLD first, and reacted in MAI vapor for various time. As shown in Figure 2(d), after the reaction with insufficient time (20 minutes at 220 °C), a strong peak of PbI$_2$ at 13.1° appeared, indicating the partly transformation of PbO to PbI$_2$. When the reaction sustained for 60 minutes, the peak of PbI$_2$ at 13.1° disappeared and the peak of MAPbI$_3$ at 13.9° appeared, indicating the complete transformation of PbI$_2$ to MAPbI$_3$.[15] Therefore, the mechanism of the reaction of transformation from PbI$_2$ to MAPbI$_3$ (Figure 3) can be mainly divided into three steps: (1) Part decomposition of MAI to HI and CH$_3$NH$_2$ at high temperature. (2) Reaction between metallic Pb in the film and HI to form PbI$_2$. (3) Transformation of PbI$_2$ to MAPbI$_3$ by reacting with the MAI gas produced by sublimation of MAI powder.

The morphology change of PbI$_2$ to MAPbI$_3$ is shown in Figure 4. It can be seen that PbI$_2$ thin film fabricated by PLD is very compact and uniform with a thickness of about 180 nm. No clear grain can be observed from the cross-sectional and plane-view of the scanning electron microscopy (SEM) images, indicating the much small-sized grains in as prepared PbI$_2$ film, which is due to the low substrate temperature of room temperature during the deposition of PLD. When the reaction is complete the grain size of MAPbI$_3$ is significantly increased, which is of about 1 μm under the reaction temperature of 180 °C and can be clearly resolved by SEM. Generally, the single grain penetrates throughout the whole film, since the grain size is much larger than the film thickness, which is only about 400 nm. Furthermore, the grains connected to each other tightly without break. The obviously increased grain size can significantly reduce the grain boundary, especially in the film thickness direction, which can effectively decrease the defect density. This might suppress the recombination of the photon-induced electron-hole pairs, increase the efficiency of the related solar cells.

We further fabricated $MAPbI_3$ thin film from $PbI_2$ film in MAI vapor under various temperatures from 160 $^oC$ to 220 $^oC$ at the same time of 20 minutes. As can be seen from the XRD pattern shown in Figure 5, diffraction peaks of $PbI_2$ can be observed at sintering temperature lower than 200 $^oC$, which was due to the incomplete reaction with MAI vapor. However, the peak intensity of diffraction peaks of $PbI_2$ decreased with increasing the sintering temperature, and disappeared when the $PbI_2$ thin film reacted in resistance furnace at 200 $^oC$. As the reaction temperature further increased, $PbI_2$ peak appeared again because $MAPbI_3$ decomposed into $PbI_2$ again. Plane-view SEM images (Figure 6(a)-(e)) of the $MAPbI_3$ thin film at various reaction temperature indicate that $MAPbI_3$ grain size increased with the rise of reaction temperature and tens of micrometer can be obtained over 200 $^oC$. We further checked the influence of reaction time on the formation of $MAPbI_3$ phase, and found that the reaction time should be prolonged with lowering the reaction temperature, as shown in Figure S1.

$MAPbI_3$ grain size distribution at different reaction temperature is presented in Figure 7(a)-(e). When we selected 160 $^oC$ as reaction temperature, grain size was mainly distributed within 1.2-1.8 μm and size of a fraction of grains was less than 1 μm still exists. When the reaction temperature increased to 180 °C, grain size distribution range was broadened to 3.5 μm and grain size was mainly distributed in 1.0-2.5 μm. All grain size was mainly distributed in 3.0-6.0 μm when we continued to increase reaction temperature to 190 °C. When the reaction temperature was 200 °C, the main grain size was mostly in the range of 15-20 μm and partly was over 20 μm. When the reaction temperature was further increased to 210 °C, grain size was mainly distributed in 10-50 μm and some grains' size even reached 60 μm. The relationship between average grain size and reaction temperature is displayed in Figure 7(f). As the reaction temperature increased to 180 °C, average grain size gradually increased, which increased drastically when the sintering temperature was above 190 °C. Thus, MAPbI$_3$ thin film with grain size over tens of micrometer can be successfully prepared using the method introduced in this work by properly selecting the optimized sintering temperature. We further conducted the research on the stability of MAPbI$_3$ thin films shown in the Figure S2. For comparison, we prepared MAPbI$_3$ thin films by our method and anti-solvent spin-coated method.[8] We put them without any encapsulation into a closed box where humidity was maintained to be above 90% and photographed them in a fixed interval of one day. Significant decomposition of the MAPbI$_3$ thin film fabricated by the anti-solvent method can be observed one day later and the film was fully discomposed two days later. However, the decomposition of the MAPbI$_3$ thin film fabricated by our method was much slower. Even after one week, the main part of the film was still kept and no decomposition can be observed. Thus, the MAPbI$_3$ thin film with large grain size prepared by our method possessed higher humidity resistance.

We further fabricated the solar cells using MAPbI$_3$ thin film prepared by this method at 180 °C whose grain size was around 1 μm. The cross-sectional SEM image of the cell is exhibited in Figure 6(f). The whole perovskite solar cell layer was composed of 100 nm thick compact TiO$_2$ layer as electron transport layer, 250 nm thick MAPbI$_3$ layer as light absorption layer, 200 nm thick hole transport layer made by Spiro-OMeTAD and 100 nm thick Ag layer as electrode. J-V curve (Figure 8) of the cell shows that short-circuit current density (J$_{sc}$) is 15.75 mA/cm$^2$, open-circuit voltage is 0.49 V, fill factor is 71.66% and conversion efficiency of the solar cell is 5.75%. It should be noted that the cell performance is still needed to be improved. Three-dimensional surface topography images of PbI$_2$ thin film prepared by PLD and MAPbI$_3$ thin film at 180 °C taken by AFM are shown in Figure 9(a) and (c). Root Mean Square Roughness (RMS) of the PbI$_2$ thin film and MAPbI$_3$ thin film were 14.9 nm and 43.1 nm, respectively. We further fabricated MAPbI$_3$ thin film by spin-coating method whose efficiency was around 15% and RMS of it was only 10.0 nm (Figure S3). The surface roughness was significantly smaller for the film fabricated by the spin-coating method indicating the better surface quality.[30] Electrochemical impedance spectroscopy (EIS) measurements were performed to further clarify the interface quality on the cell performance, as shown in Figure 10. The equivalent circuit is shown as inset to fit the EIS data. Series resistance R$_s$ accounts for the resistance of conductive substrates and wire electrode, the contact resistance R$_{sc}$ is related to interface contacts with perovskite thin film and the recombination resistance

$R_{rec}$ is the resistance of the perovskite layer and its interface.[26,31,32] In the Nyquist plots, the high frequency part (left arc) and the low frequency part (right arc) are related to contact resistance ($R_{sc}$) and recombination resistance ($R_{rec}$) respectively.[33] From the figure we can see that the arcs of two methods in the high frequency part have nearly similar radian and the radius of the arc of the anti-solvent method is obviously bigger than that of the arc in this work which means lower $R_{rec}$ in this work. Lower $R_{rec}$ indicates that more recombination occurs at the $TiO_2$/$MAPbI_3$/HTM interface due to seriously rough surface of the $MAPbI_3$ layer.[34,35] Therefore $MAPbI_3$ thin film prepared from the high-temperature reaction of $PbI_2$ thin film with MAI vapor had much larger grain size which had much less grain boundaries, the increased surface roughness might cause serious wettability problem and charge carrier recombination when the top hole transport layer of Spiro-OMeTAD. The contact angle between HTM solution and $MAPbI_3$ thin films in our work is shown in the Figure S4(a). For comparison, the contact angle between HTM solution and the $MAPbI_3$ thin film fabricated by the anti-solvent method with around 15% efficiency is also shown in Figure S4(b). It's easy to see that the contact angle in Figure S4(a) is evidently larger than that in Figure S4(b) which means a bad contact for HTM with $MAPbI_3$ thin film due to increasing of surface roughness, resulting in a poor performance of device. And also due to the growing grains during the high temperature reaction process, the bottom interface between $TiO_2$ and $MAPbI_3$ layer might also be significantly deteriorated. This can be clearly seen from Figure 6(f) that some cracks can be observed between $TiO_2$ and $MAPbI_3$ layer. Thus, the defect density at the top and bottom interfaces of $MAPbI_3$ layer was obviously increased, leading to the strongly decreased efficiency. As can be seen from Figure 8, the performance of solar cells is low, which depends on collection of charges and recombination in perovskite thin films interface.[30]

Thus, further improvement of the smoothness of the top surface of MAPbI$_3$ layer and compactness between TiO$_2$ layer and MAPbI$_3$ layer might significantly improve the efficiency and the photovoltaic performance of the solar cell, which is being under investigation.

**Conclusion**

In summary, PbI$_2$ thin film prepared by PLD has been successfully transformed to pure phase MAPbI$_3$ thin film by the high temperature reaction in MAI vapor. The grain size of MAPbI$_3$ layer can be in micrometer size by properly selecting reacting temperature of above 160 °C, and can be further increased to tens of micrometer size with further increasing the reaction temperature above 190 °C. Due to that the grain size was in micrometer size, which is much larger than the thickness, most of the grains can penetrate through the whole film without grain boundaries. Thus, the recombination of the electron-hole pairs during the transport through the MAPbI$_3$ layer can be significantly suppressed. The MAPbI$_3$ thin film shows better humidity resistance than that prepared by the spin-coated method. The planar structured solar cells using the MAPbI$_3$ layer prepared by this method at 180 °C were prepared and the short-circuit current density ($J_{sc}$) is 15.75 mA/cm$^2$, open-circuit voltage is 0.49 V, fill factor is 71.66% and conversion efficiency of the solar cell is 5.75%. AFM images showed that the surface roughness of MAPbI$_3$ layer may influence the wettability of

Spiro-OMeTAD on it, and deteriorate the interface between TiO$_2$ and Spiro-OMeTAD, leading to the smaller open-circuit voltage and short-circuit current. This work provides an efficient method for the preparation of MAPbI$_3$ thin film for the further fundamental and application researches.

**Materials**

All the chemicals and reagents in the experiments can be available commercially. Glass coated Fluorine-doped tin oxide (7 Ω/sq) was purchased from Nippon Sheet Glass Company. Lead iodide (AR, >98%), acetonitrile and Hydroiodic acid (AR, 45wt% in water) was obtained from Sinopharm Chemical Reagent Co. Ltd. Tetrabutyl titanate, absolute ethyl alcohol, methylamine (AR), isopropyl alcohol, chlorobenzene were bought from aladdin. Spiro-OMeTAD was received from Feiming technology Co. Ltd. Lithium bis(trifl uoromethanesulfonyl)imide (Li-TFSI) and tributyl phosphate (TBP) were acquired from Xi'an Polymer Light Technology Co. Cobalt bis (trifl uoromethanesulfonyl) imide (Co-TFSI) was gotten from Shanghai Materwin new materials Co. Ltd. Argent grain (99.999%) was supplied by Beijing PuRui new materials Co. CH$_3$NH$_3$I (MAI) powders were synthesized with Hydroiodic acid and methylamine.[31]

**Methods**

Initially, FTO-glass was chosen as substrate and cut into the size we wanted, then was etched using Zn powder and diluted HCl acid and cleaned with detergent water, deionized water and absolute ethyl alcohol by ultrasonic. Compact TiO$_2$ film whose thickness was about 80 nm was deposited on FTO substrate by spin-coating the tetrabutyl titanate solution (18% in absolute ethyl alcohol) at 5000 rpm for 30 seconds and annealing for 30 minutes at 450 °C.

The following step was the preparation of $PbI_2$ thin film by PLD. The target of PLD was prepared using $PbI_2$ powder. Prepared compact $TiO_2$ substrate and target were put into the chamber which evacuated down to less than $10^{-3}$ Pa. The wavelength of laser from the excimer laser made by Coherent Inc. was 248 nm. We set the laser energy to be 100 mJ and pulse repetition rate to be 5 Hz. The number of pulses we closed was 400 to make the thickness of the final $MAPbI_3$ thin film be about 400 nm. The whole process was performed at room temperature without shielding gas.

We put the $PbI_2$ film and the abundant MAI powder into a ceramic container in air and the $PbI_2$ film was above the MAI powder. The ceramic container was put into a resistance furnace and annealed at different temperature from 160 °C to 220 °C for different time without any protecting gas. When the heating time was over, the ceramic container was taken out immediately and cooled to room temperature. Then, we removed the redundant MAI by spin-coating the isopropyl alcohol at a low rate.

Finally, a hole transport materials layer with a thickness of about 250 nm was deposited by spin-coating the hole transporting material (HTM) solution mixed of 288 μL TBP, 175 μL Li-TFSI solution (520 mg Li-TSFI in 1 ml acetonitrile), 290 μL Co-TFSI solution (300 mg Co-TSFI in 1 ml acetonitrile) and 20 mL chlorobenzene on the $MAPbI_3$ thin film at 5000 rpm for 30 seconds. After one night standing, 100 nm thick Ag film was deposited on the HTM film as back electrode by thermal evaporation at 0.2-0.3 Å/s rate in the chamber under the pressure of $10^{-4}$ Pa.

**Measurement**

In this experiment, we measured XRD patterns by an X-Ray diffractometer (Rigaku Smartlab3) which used Cu Kα as the radiation source to analyze the constituent in film. Images of cross-section and surface of samples were obtained by a SEM (FEI Inspect F50). The PL spectra were obtained by a Raman spectrometer made by Horiba Jobin Yvon using a laser of 325 nm wavelength to investigate the band gap of the samples. AFM images were obtained by a BioScope Resolve™. Current-voltage (I-V) was measured by Keithley 2400 under Newport Oriel 91.192 simulated illumination (AM1.5, 100 mW/cm$^2$). EIS was measured with electrochemical workstation produced by Shanghai ChenHua instruments Co.

## Acknowledgments

This work is supported by the National Natural Science Foundation of China (51471085, 11674055), the Natural Science Foundation of Jiangsu Province of China (BK20151400), and the open research fund of Key Laboratory of MEMS of Ministry of Education, Southeast University.

## Author contributions

Q.X. conceived the experiment idea and designed the research. S.D. and Q.L. discussed and analyzed the results. H.Z. carried out the experiment and did most measurement. M.T. implement fabrication of solar cells, B.G. designed ceramic container, W.C. implemented fabrication of PbI2 thin films by PLD. H.Z., Q.X., and S.D. wrote the paper.

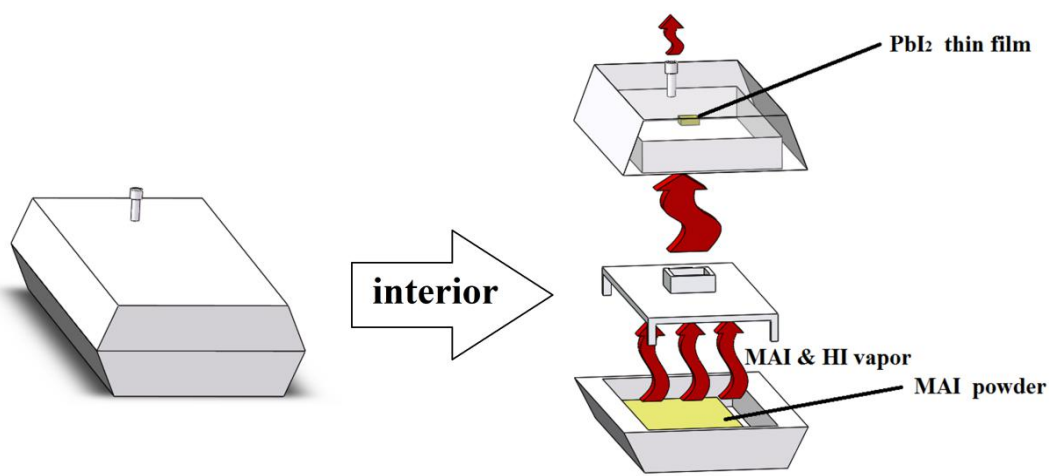

**Figure 1.** Schematic representation of the ceramic container where MAI powder and PbI$_2$ thin film were put.

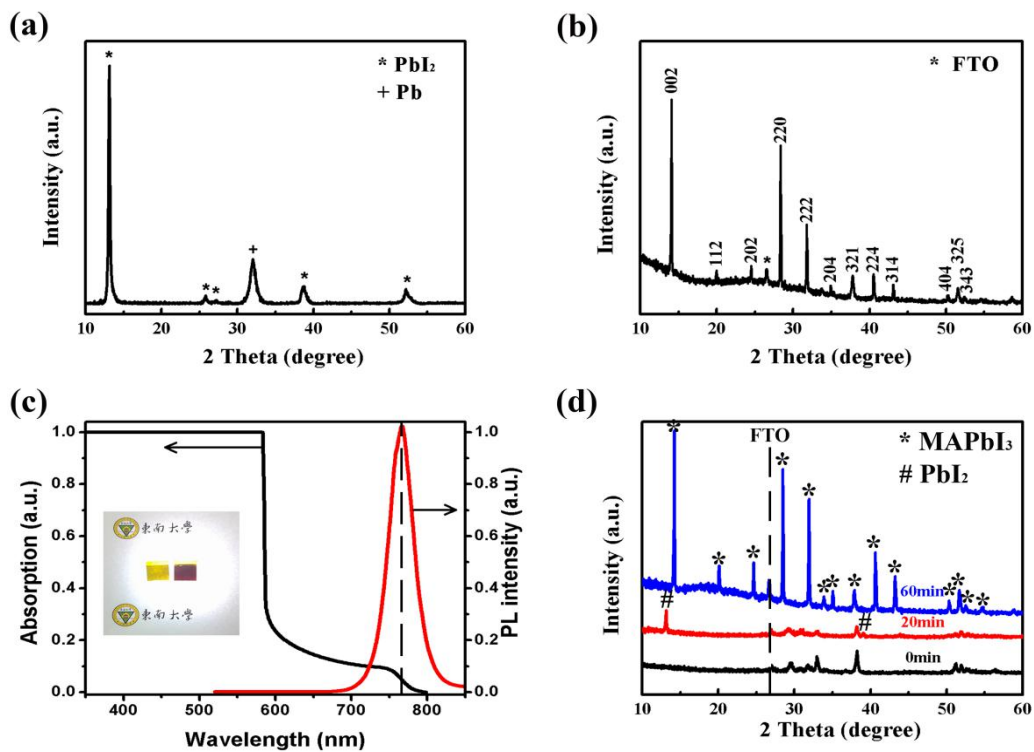

**Figure 2.** XRD patterns of (a) PbI$_2$ film grown by PLD and (b) MAPbI$_3$ thin film prepared at 180 °C for 25 min. (c) The PL and absorption spectra of the MAPbI$_3$ thin film with photos of samples (the left is the PbI$_2$ thin film and the right is the corresponding fabricated MAPbI$_3$ thin film). (d) XRD pattern of PbO thin films reacted in resistance furnace at 220 °C for 0 minutes, 20 minutes and 60 minutes, respectively.

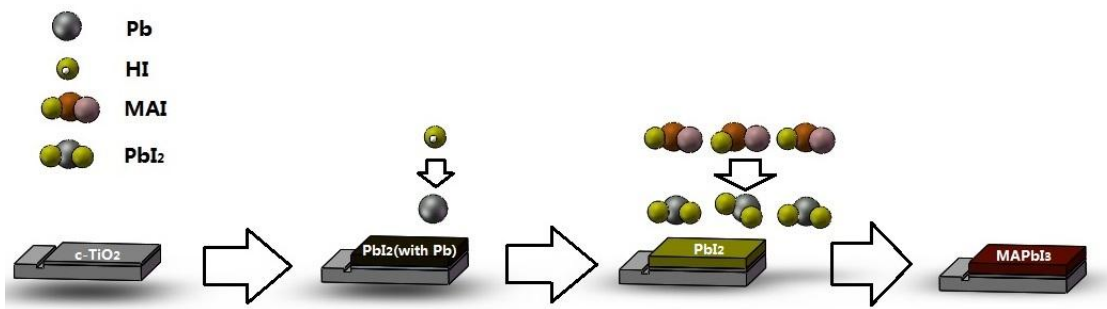

**Figure 3.** Diagrammatic sketch of the transformation from PbI$_2$ thin film to MAPbI$_3$ thin film with MAI vapor in resistance furnace at high temperature.

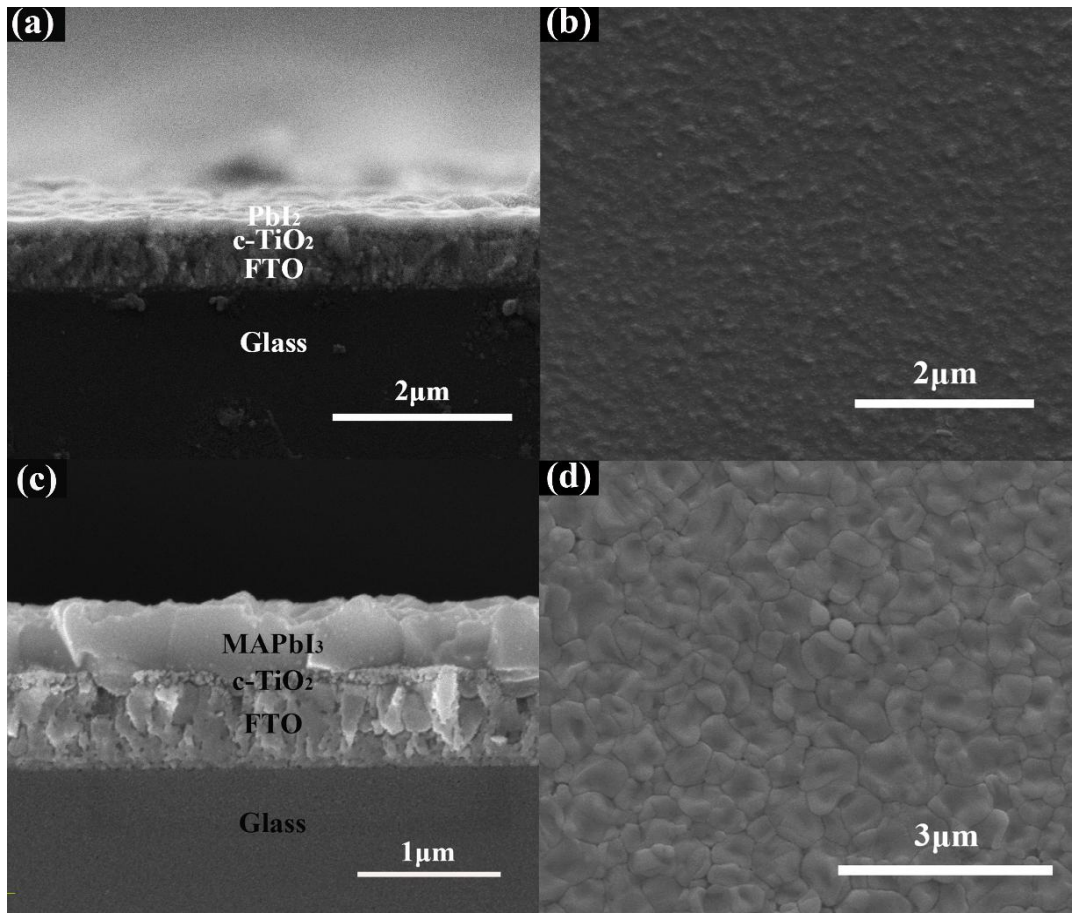

**Figure 4.** (a) Cross-sectional and (b) Plane-view SEM images of PbI$_2$ thin film fabricated by PLD. (c) Cross-sectional and (d) Plane-view SEM images of prepared MAPbI$_3$ thin film reacted at 180 °C for 25 minutes.

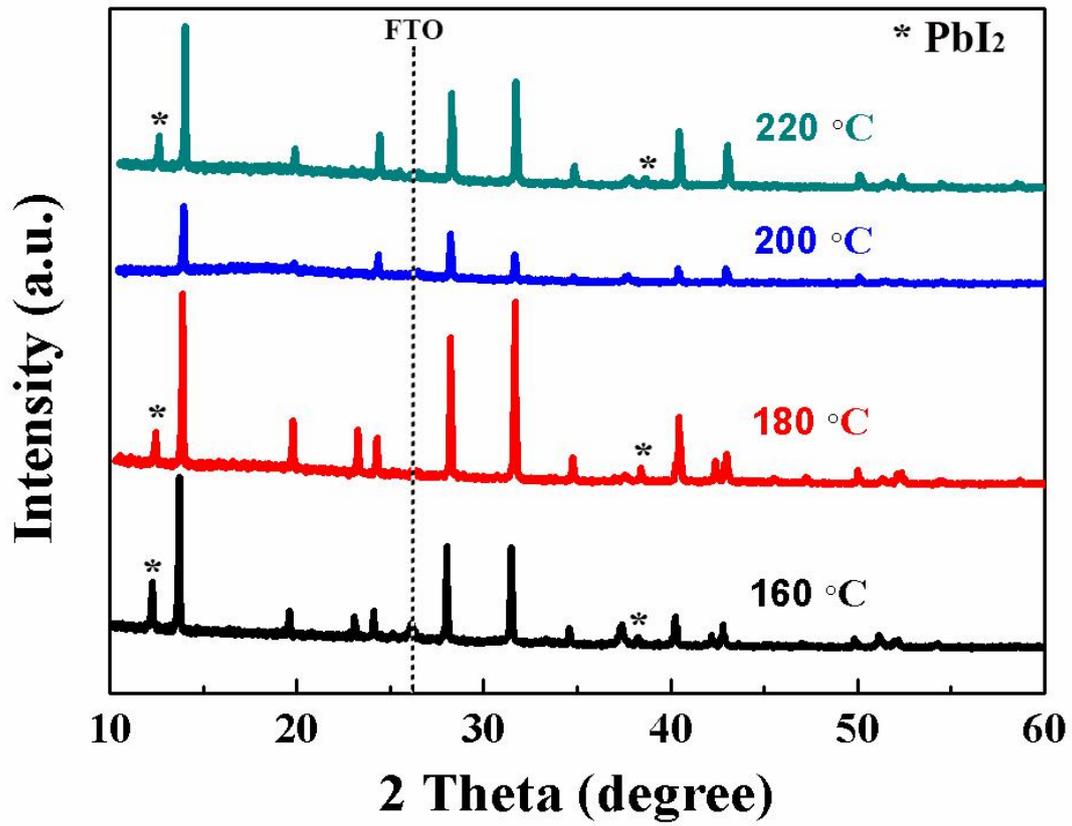

**Figure 5.** XRD pattern of PbI$_2$ thin films reacted in resistance furnace in MAI atmosphere at different temperature from 160 °C to 220 °C for 20 minutes.

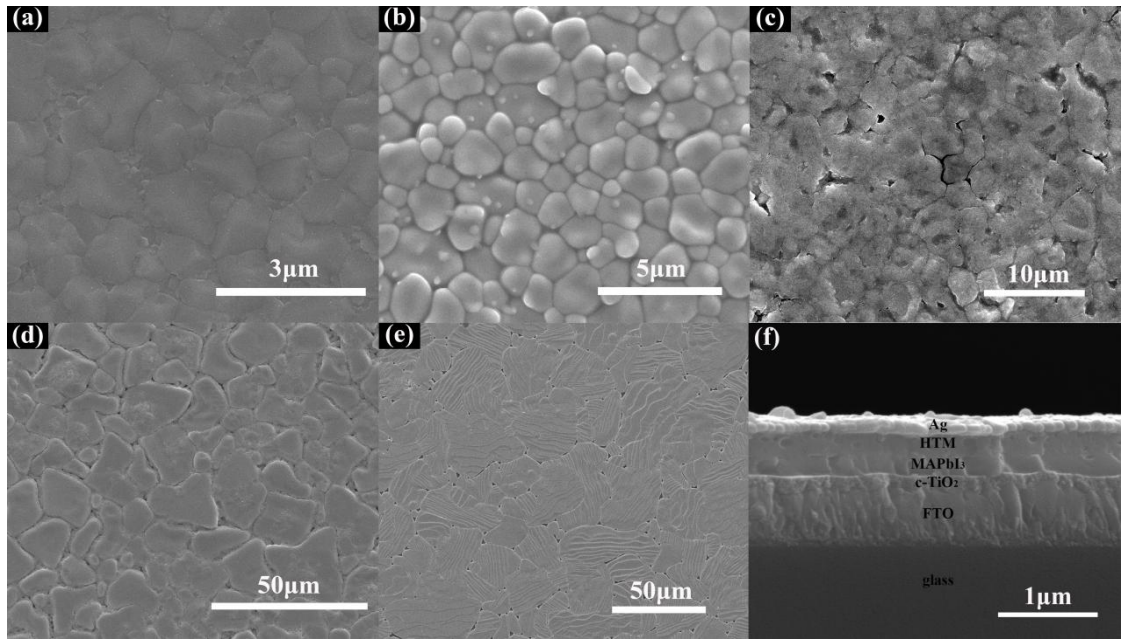

**Figure 6.** Top-view SEM images of the MAPbI₃ thin films prepared at (a) 160 °C for 30 min, (b) 180 °C for 25 min, (c) 190 °C for 22 min, (d) 200 °C for 20min and (e) 210 °C for 15 min. (f) Cross-sectional SEM image of the prepared device when reaction temperature of MAPbI₃ layer was 180 °C.

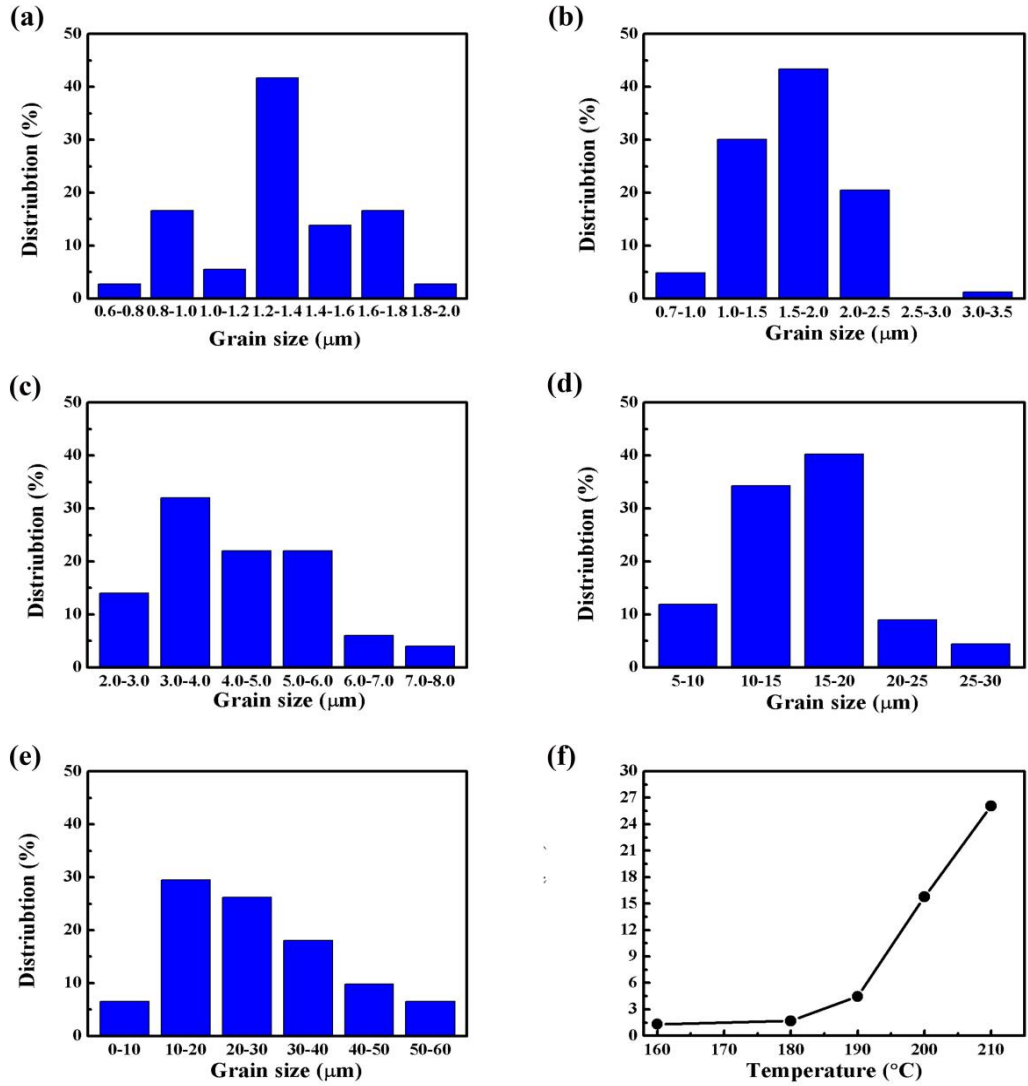

**Figure 7.** Histograms of MAPbI$_3$ grain size distribution prepared at (a) 160 °C, (b) 180 °C, (c) 190 °C, (d) 200 °C and (e) 210 °C, for 20 minutes. (f) The relationship of the average MAPbI$_3$ grain size with the preparation temperature.

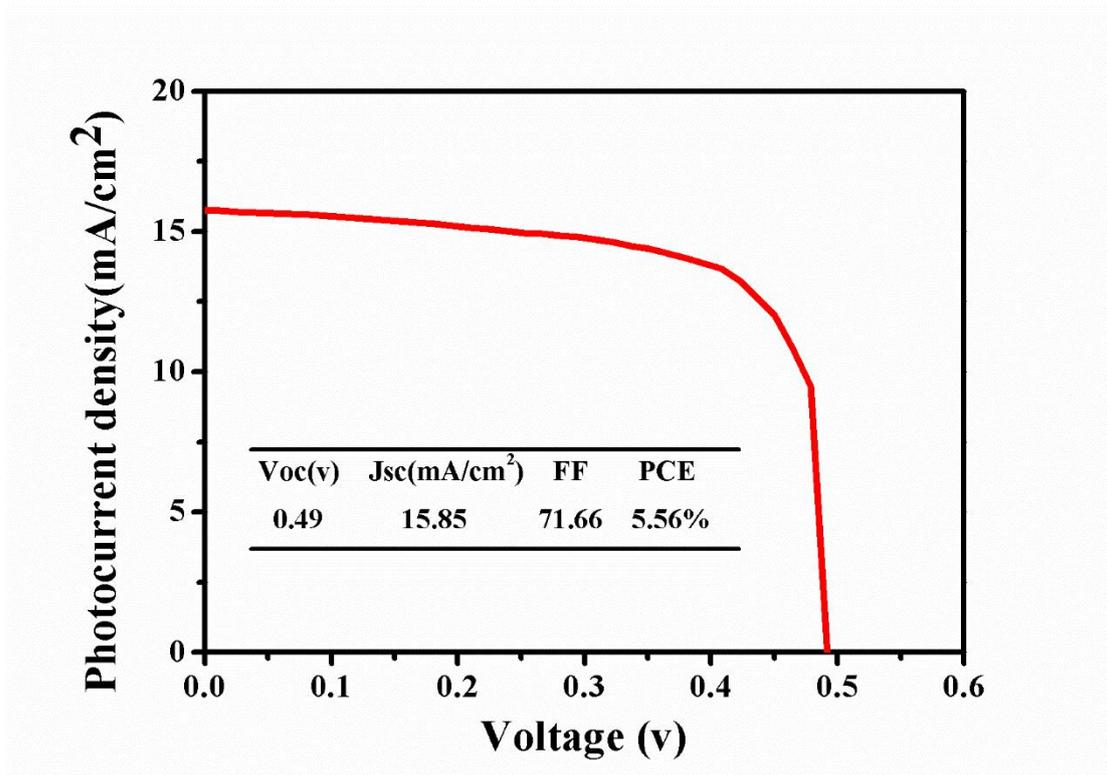

**Figure 8.** J-V curve for the prepared plane structured solar cell, using MAPbI$_3$ thin film reacted at 180 °C for 25 minutes.

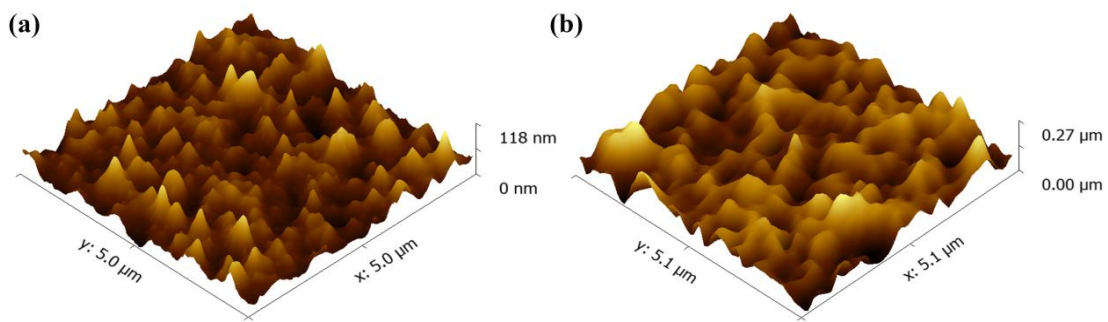

**Figure 9.** Three-dimensional AFM images of (a) PbI$_2$ thin film fabricated by PLD and (b) prepared MAPbI$_3$ thin film at 180 °C for 25 minutes.

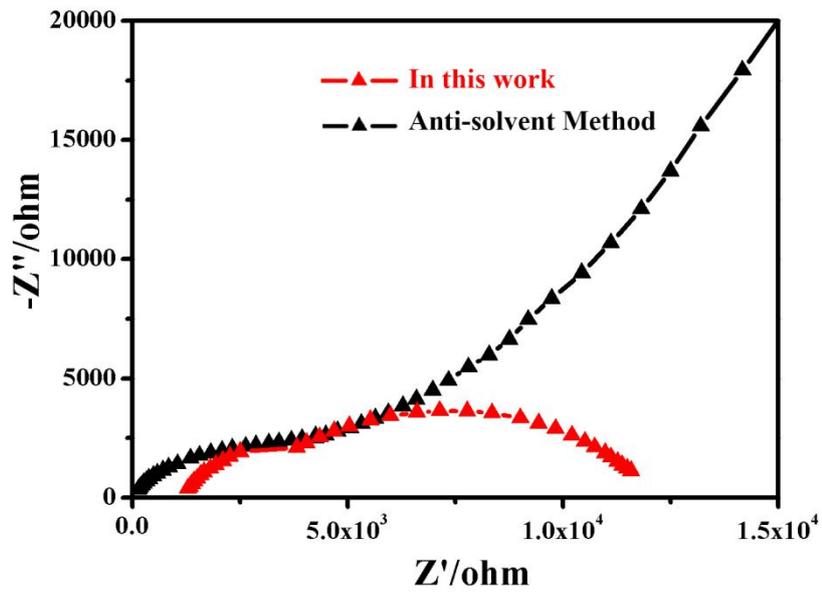

**Figure 10.** Nyquist plot of PSC devices prepared by this method (red) and anti-solvent method (black) in the dark.

# Supplemental material

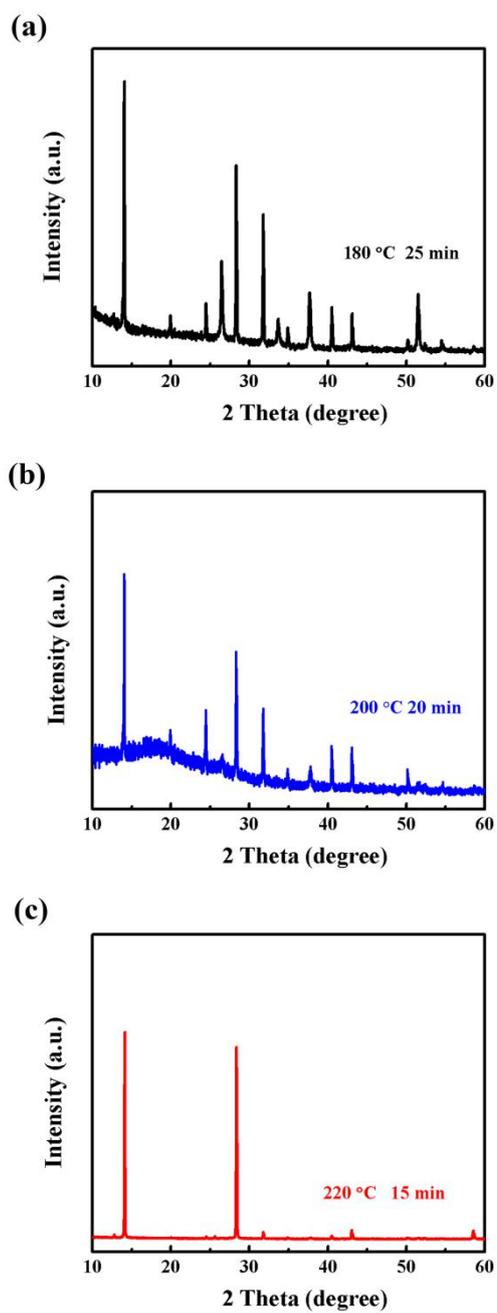

**Figure S1** XRD pattern of MAPbI$_3$ thin films prepared (a) at 180 °C for 30 min, (b) at 200 °C for 20 min, (c) at 220 °C for 15 min.

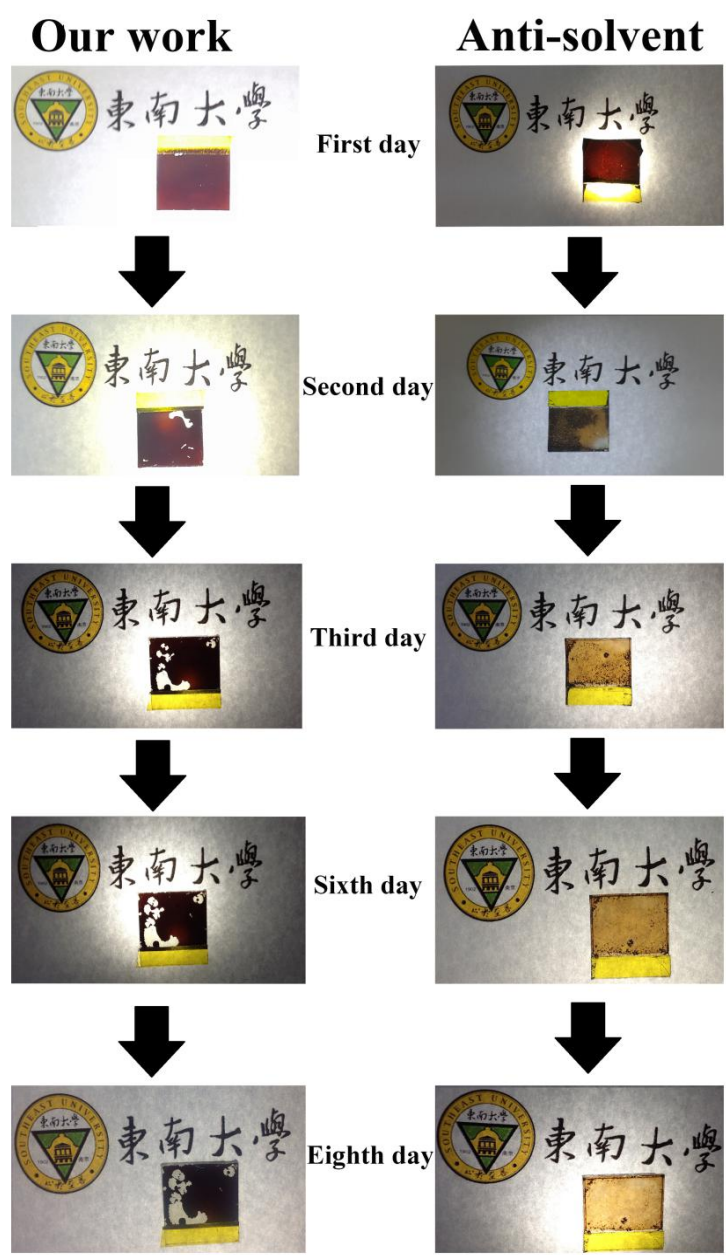

**Figure S2** Photos of MAPbI$_3$ thin films prepared by our method and the anti-solvent method under the humidity of above 90% for various days.

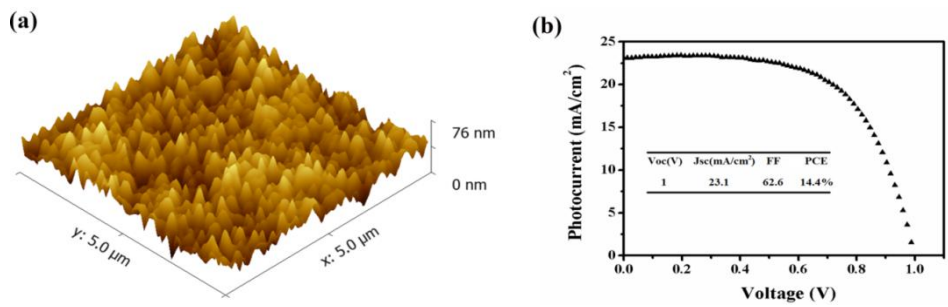

**Figure S3** (a) Three-dimensional AFM image, the surface roughness (RMS value) is 10 nm and (b) J-V curve for the solar cell using MAPbI$_3$ film prepared by spin coating.

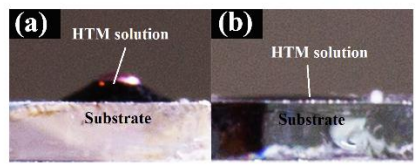

**Figure S4** Images of contact angles between HTM chlorobenzene solution and MAPbI$_3$ thin films prepared by (a) our method and (b) the anti-solvent method.